# Fault tolerant leader election in distributed systems

Marius Rafailescu

The Faculty of Automatic Control and Computers, POLITEHNICA University, Bucharest

*ABSTRACT*

*There are many distributed systems which use a leader in their logic. When such systems need to be fault tolerant and the current leader suffers a technical problem, it is necesary to apply a special algorithm in order to choose a new leader. In this paper I present a new fault tolerant algorithm which elects a new leader based on a random roulette wheel selection.*

## 1. INTRODUCTION

Leader election is a common part of distributed systems. For instance, we may consider a distributed database where all the nodes need to reach the commit consensus. Most of the algorithms designed for this consensus problem use a node with a special role, the leader. In Paxos, the author does not present a specific leader election method [1] and this is why I tried to define a new fault tolerant algorithm.

In the past, several solutions where developed for synchronous and asynchronous distributed systems. Garcia-Molina has one important result for each system [2]. Later on, many other authors tried to extend/optimize his work [3], [4].

Depending on the network topology, other algorithms have been presented until today, the Ring Election Algorithm [5] being one example.

In Raft, a recent consensus protocol for replicated state machines, is specified an algorithm for leader election, which is simple and elects a new leader based on randomized timeouts - the node with the lowest timeout becomes candidate after the timeout passes and becomes leader when the other nodes vote it, in majority [6]. Here, when two nodes have (almost) the same timeout, appears the situation of split votes.

Besides consensus problem, leader election algorithm is a basic component in many protocols, especially in environments where special tasks in the network need to be done by a specific node or where self-organization of network is necessary, such as wireless ad-hoc networks.

In this paper, the proposed algorithm tries to minimize the split votes using a stronger candidate position condition and gives a chance to all nodes because the final leader can differ from the candidate (more exactly, it can differ from the round of vote coordinator, as described next in section 2).

The work has been funded by the Sectoral Operational Programme Human Resources Development 2007-2013 of the Ministry of European Funds through the Financial Agreement POSDRU 187/1.5/S/155420.

## 2. LEADER ELECTION ALGORITHM

In fault tolerant systems which use a node with a special role, the leader, there may be cases when this node has technical problems and it is not able to work properly. In this cases, we have to apply a clear procedure in order to choose a new one. The algorithm elects randomly a new leader between the remaining running nodes and only one node must be chosen based on some random condition. Because such condition may be hard to satisfy, in order to give almost the same chances for all the nodes, we can use a random roulette wheel selection approach, for instance. The algorithm is explained below:

1) Each node uses a finite local mechanism which collects all necessary informations from other nodes in order to use eventually a random roulette wheel selection; this local logic has to ensure that the node is eligible to run the roulette wheel only by satisfying the launch condition, which refers, for instance, to generating a random number in (0, 1) interval, but greater than chosen threshold, let's say 0.85; when the condition is matched, the node becomes a candidate and sends to all other nodes a special message with this number;

2) When one node receives the number generated by other node, it has to vote for sender node if it has not voted before for a bigger value in current round of vote; moreover, it sends his greatest random generated number in his response;

3) The node which receives all the positive votes will be the coordinator in order to run the random roulette wheel selection using all the numbers received from other nodes - the winner will be the new leader;

4) After the leader is chosen, the coordinator will send it a message and after receiving it, the new leader will broadcast a special/heartbeat message.

### 2.1. CASE STUDIES

Let's consider a cluster with 5 nodes and 2 cases: the first with only one candidate and the second with two candidates for the coordinator position.

### 2.1.1. FIRST CASE

First of all, we have the simplest case when only one node satisfies the launch condition. In the figure 1, it is ilustrated this case where node A is the candidate for the coordinator state.

The figure shows the messages sent from A to others nodes and we will consider that every node responds right after the message from A is received. Because A is the only candidate, it will receive positive messages from all other nodes; combining all the numbers received from them it will choose node C as the new leader. In the end, node C will send heartbeat messages to announce its new role.

### 2.1.2. SECOND CASE

In the second case study, we have two nodes which satisfied the launch condition nearly in the same time. In the figure 2, it is ilustrated this case where node A and node B are the candidates for the coordinator state.

The work has been funded by the Sectoral Operational Programme Human Resources Development 2007-2013 of the Ministry of European Funds through the Financial Agreement POSDRU 187/1.5/S/155420.

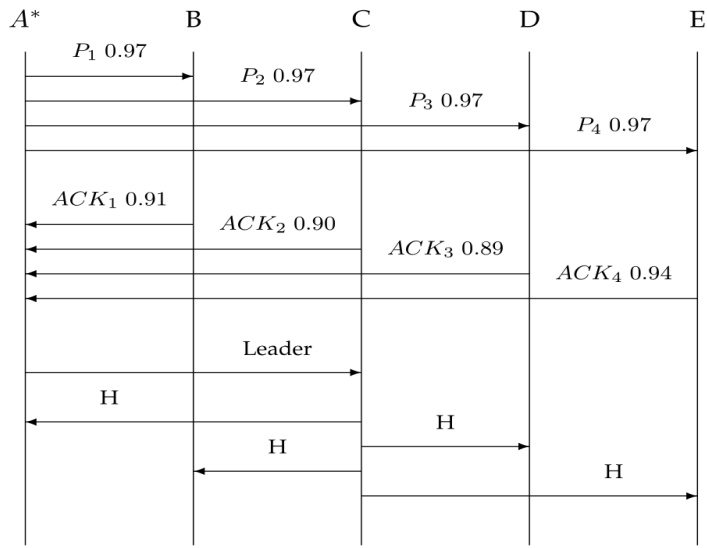

Fig. 1. Case Study 1

The figure shows that node A sends the first two messages to nodes D and E, after which it receives two positive votes. Because node A initially votes for itself, in this moment it has the majority of votes, but it waits to see the responses from other nodes.

In the next two messages, node B sends its first messages to nodes C and D. Because its greatest random generated number is greater that A's number, it will collect two positive responses (node D votes for B after it voted for A in the first couple of messages).

Next, proposal from node A reach node C, but the latter will give a negative response because it voted later with B's greater value.

Node B sends its proposal message to node A which will vote because its number its lower. After that, B gets one more positive answer and A one negative. B becomes the coordinator and after it runs the random roulette wheel, node C is chosen as the new leader. In the end, node C will send heartbeat messages to announce its new role.

## 2.2. ADDITIONAL SPECIFICATIONS

One important result of previous work on consensus algorithms is that no completely asynchronous consensus protocol can tolerate even a single unannounced process death [7]. That is why it is necessary to use timeout-based methods in order to ensure correct termination for such an algorithm.

The work has been funded by the Sectoral Operational Programme Human Resources Development 2007-2013 of the Ministry of European Funds through the Financial Agreement POSDRU 187/1.5/S/155420.

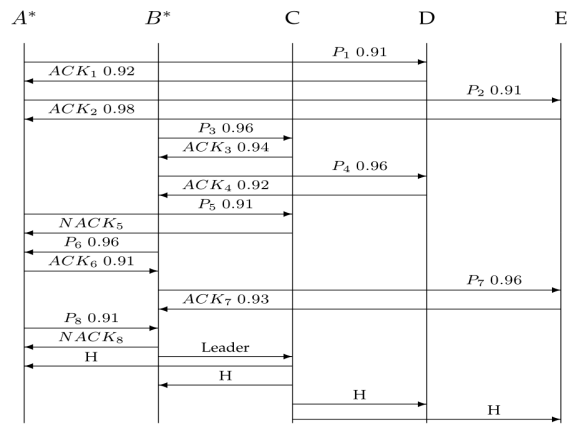

Fig. 2. Case Study 2

There are some special cases which need additional specifications:

1) If we consider the solution based on random number generation, we have to admit that multiple nodes might generate nearly in the same time some numbers greater than 0.85. The probability for one node to generate a random number greater than 0.85 is 0.15. To limit this probability, we could take into consideration changing this condition; we can redefine the rule such as one node need to generate one number greater than threshold multiple times sequentially; so the probability will be 0.0225 for two numbers and 0.003375 for three numbers;

Also it is necessary to specify that every node which managed to satisfy the condition, votes first of all for itself and it will vote for other node only if the number received is greater than its own number; moreover, when such a node receives a negative response, it can revert its state as long as this means that its number is lower;

Furthermore, it could be a problem if two nodes generate the same big number. The solution consists in generating random numbers with many decimal places (6 would be sufficient). However, if this happens and no coordinator can be selected, the algorithm will continue with another round of vote;

2) One node becomes coordinator by waiting responses from all the nodes, using a timeout - when all the nodes respond or the timeout elapse, the candidate verifies the votes and it will become coordinator if it has the majority of them and, eventually, it will run the roulette wheel with all numbers from other nodes;

3) If one node obtains the majority of votes and becomes coordinator, but after that receives a proposal with a greater number, it will send a negative response with a special flag saying that it is already coordinator in the current round of vote;

4) If one node which satisfies the launch condition has technical problems, it will stop until the problem is resolved, but the algorithm will continue through a new round of voting;

5) If one node does not generate a value bigger than threshold, then it will continue to generate numbers until it gets a proper number or until it receives a proposal message from other node;

The work has been funded by the Sectoral Operational Programme Human Resources Development 2007-2013 of the Ministry of European Funds through the Financial Agreement POSDRU 187/1.5/S/155420.

6) If the coordinator suffers a problem, the algorithm will continue through a new round of vote;

7) If the node which is chosen to be the new leader suffers a problem right before the coordinator announce it, the coordinator can choose another leader if it does not receive a heartbeat message in a short time;

8) Depending on the system in which the election is used, there might be necessary some additonal steps for ensuring consistency between all nodes;

9) Every node knows anytime how many nodes are up. If at one moment of time, the majority can not be satisfied, the system will block until some nodes are recovered.

The modified algorithm is the following:

1) Each node generates random numbers in (0, 1) interval. If three consecutive generated numbers are greater than a chosen threshold, then it will send to other nodes the greatest generated number and votes for itself;

2) When one node receives the number generated by other node, it has to vote for sender node if it has not voted before for a bigger value in current round of vote; moreover, it sends his greatest random generated number in his response;

3) The node which receives all the positive votes will be the coordinator in order to run the random roulette wheel selection using all the numbers received from other nodes; the winner will be the new leader;

4) When one node receives a negative vote response, it can invalidate his state;

5) After the leader is chosen, the coordinator will send it a message and after receiving it, the leader will broadcast a special/heartbeat message.

To minimize the time a node must wait in order to become coordinator, we may consider that the roulette wheel selection can be started right after the candidate collects the majority of votes (half + 1). This aproach implies a change in every node behavior because it is mandatory to vote only one time in a round, for the first candidate node. In this way, a single node will be elected as coordinator.

As we can see, we identify multiple states a node can be in:

1) Basic mode when a node simply votes for others;

2) Candidate mode when a node already satisfied the launch condition;

3) Coordinator mode when a node received the majority of votes;

4) Leader, final state;

Also, we can see the types of messages which are send between nodes:

1) Proposal message;

2) Positive vote message;

3) Negative vote message;

4) Announcement message;

5) Heartbeat message (may be combined with consistency check message);

The work has been funded by the Sectoral Operational Programme Human Resources Development 2007-2013 of the Ministry of European Funds through the Financial Agreement POSDRU 187/1.5/S/155420.

## 2.3. NUMBER OF MESSAGES

Having $v$ nodes, there are in total $2v - 1$ messages (not taking into consideration final heartbeat messages):

1) $v - 1$ messages from the node which generated big numbers;
2) $v - 1$ responses from other nodes;
3) message from coordinator to leader.

## 2.4. ANALYSIS

One delicate discussion is about reducing split vote frequency and choosing a single coordinator, as explained above. Taking into consideration the mechanism based on randomly generated numbers which are bigger than a chosen threshold, 0.85, it is necessary to analyze which is the probability as such multiple nodes satisfy the condition.

To experiment, in 1000 iterations there were generated random numbers, using a counter until the condition is satisfied.

With only one single number generated, the probability is very high as almost all nodes generated numbers greater than threshold.

Testing the condition based on two consecutive numbers showed that the probability is also high. In the first 100 ms, over 80% of nodes managed to do that.

With three consecutive numbers, there were better results; only 22% of nodes generated big numbers in the first 100 ms, with an average of 385 ms. Using a threshold of 0.9, the probability dropped to 8% for the first 100 ms, but this time the average was 1.3 seconds.

Furthermore, considering four consecutive numbers, satisfying condition would took in average over 2 seconds, which is not acceptable.

As a result, suggestion consists in choosing the condition based on three consecutive numbers.

## 2.5. ELECTION PERFORMANCE

Testing the proposed new algorithm in its unoptimized way, in which the candidate node needs to wait for all responses from other nodes, showed that a new leader is chosen in an average time of 210 ms, with a minimum of 64 ms and maximum of 851 ms. The histogram results are presented in figure 3.

The work has been funded by the Sectoral Operational Programme Human Resources Development 2007-2013 of the Ministry of European Funds through the Financial Agreement POSDRU 187/1.5/S/155420.

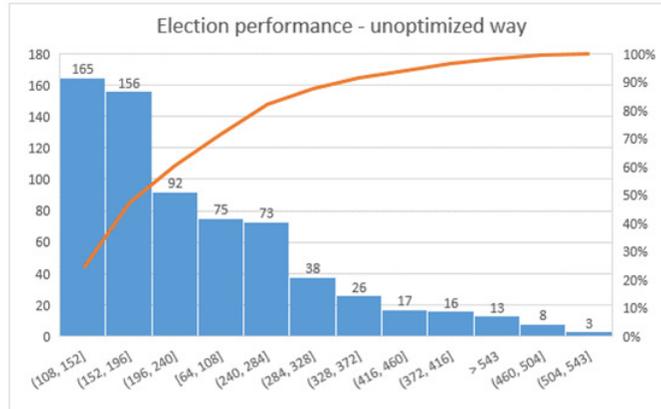

Fig. 3: Election performance – unoptimized way

Taking into consideration the optimized form of the algorithm as mentioned in section 2.2, the average time was 191 ms, histogram being shown in figure 4.

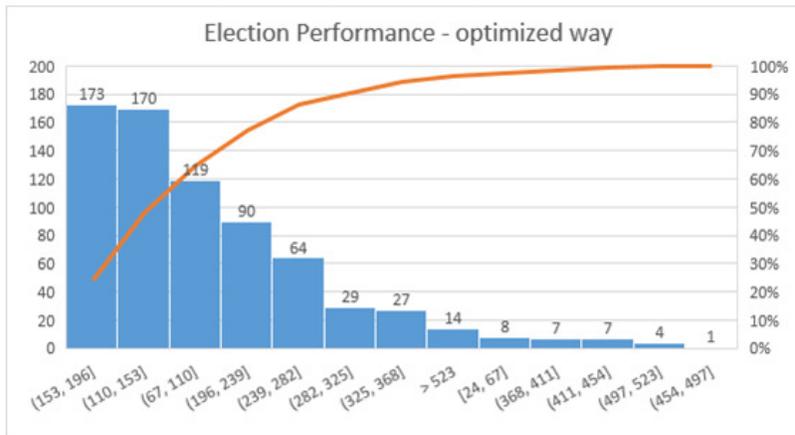

Fig. 4: Election performance – optimized way

The result is slightly better, but in both ways, in 90% of cases, the election finished in about 325 ms at most. However, the optimization should have a greater impact in bigger networks.

It is important to mention that these results are mostly influenced by the time in which the fastest node manages to satisfy the launch condition.

The tests were made using 5 nodes running on distinct virtual machines, having at least 500 iterations and using 0.85 as the trigger condition threshold.

Regarding multiple candidates possibility, in less than 10% of cases there was a split vote, as shown in figure 5.

The work has been funded by the Sectoral Operational Programme Human Resources Development 2007-2013 of the Ministry of European Funds through the Financial Agreement POSDRU 187/1.5/S/155420.

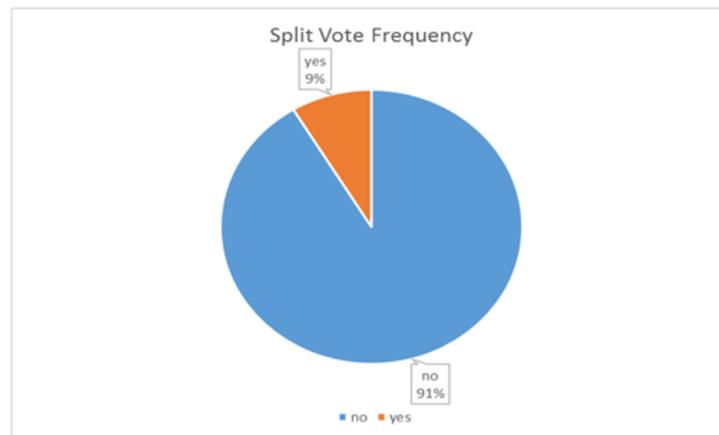

Fig. 5: Split vote frequency

# 3. CONCLUSION

The new algorithm presented in this paper is quite simple, easy to understand and time efficient. Its role is to choose a new leader from a set of nodes in a fault tolerant manner. This is done using a random selection wheel selection and is built in such a way to give a chance to every participant node.

## REFERENCES

[1] J. Gray and L. Lamport. Consensum on transaction commit. 2004.

[2] H. Garcia-Molina. Elections in a distributed computing system. IEEE Transactions on Computers, (1):48– 59, 1982.

[3] S. Basu. An efficient approach of election algorithm in distributed systems. Indian Journal of Computer Science and Engineering (IJCSE), 2(1):16–21, 2011.

[4] S.-H. Park. A stable election protocol based on an unreliable failure detector in distributed systems. Proceedings of IEEE Eighth International Conference on Information Technology: New Generations, pages 976– 984, 2011.

[5] A. Silberschatz, P. B. Galvin, and G. Gagne. Operating Systems Concepts. John Wiley & Sons. Inc, 7 edition, 2005.

[6] D. Ongaro and J. Ousterhout. In search of an understandable consensus algorithm (extended version). 2014.

[7] M. J. Fischer, N. A. Lynch, and M. S. Paterson. Impossibility of distributed consensus with one faulty process. Journal of the Association for Computing Machinery, 32(2):398–407, 1985.

## AUTHORS

**Marius Rafailescu** is a Ph.D. candidate at the Department of Computer Science at The "Politehnica" University from Bucharest. His M.S. and B.S were received also from The "Politehnica" University from Bucharest. His main research interests are transactional processing in databases and distributed systems.